\documentclass{elsart3}

 \usepackage{graphicx}

\usepackage{amssymb}
\newcommand{\mib}[1]{\mbox{\boldmath $#1$}} 
  
\begin{document}

\begin{frontmatter}

\title{
Monte Carlo study of half--magnetization plateau
and magnetic phase diagram
in pyrochlore antiferromagnetic Heisenberg model
}

\author[aff1]{Yukitoshi Motome\corauthref{cor1}} 
\ead{motome@riken.jp}
\ead[url]{http://www.riken.jp/lab-www/cond-mat-theory/\\
motome/index-e.html}
\corauth[cor1]{}
\author[aff2]{Karlo Penc}
\author[aff3,aff4]{Nic Shannon}
\address[aff1]{RIKEN (The Institute of Physical and Chemical Research), 
2-1 Hirosawa, Wako, Saitama 351-0198, Japan} 
\address[aff2]{Research Institute for Solid State Physics and Optics, 
H-1525 Budapest, P.O.B. 49, Hungary}
\address[aff3]{Department of Advanced Materials Science,
Graduate School of Frontier Sciences, University of Tokyo, 5-1-5, 
Kashiwanoha, Kashiwa, Chiba 277-8851, Japan}
\address[aff4]{CREST, Japan Science and Technology Agency, Kawaguchi 332-0012, Japan}




\begin{abstract}
The antiferromagnetic Heisenberg model on a pyrochlore lattice
under external magnetic field is studied by classical Monte Carlo simulation.
The model includes bilinear and biquadratic interactions;
the latter effectively describes the coupling to lattice distortions.
The magnetization process shows a half--magnetization plateau 
at low temperatures,
accompanied with strong suppression of the magnetic susceptibility.
Temperature dependence of the plateau behavior is clarified.
Finite--temperature phase diagram under the magnetic field is determined.
The results are compared with recent experimental results 
in chromium spinel oxides.
\end{abstract}

\end{frontmatter}

\section{Introduction}\label{sec1}

Geometrically frustrated magnetism is one of the long--standing issues
in condensed--matter physics.
Geometrical frustration leads to nearly--degenerate ground--state manifolds
of a large number of different spin configurations.
The degeneracy may yield nontrivial phenomena, 
such as spin liquid states and glassy states
\cite{Diep1994,Liebmann1986}.
Because of the small energy scale of the nearly--degenerate manifolds,
a tiny perturbation can be relevant to lift the degeneracy in these systems.
For example, quantum or thermal fluctuations can
give rise to non--trivial phase transitions
in applied magnetic field, leading to a plateau or jump
in the magnetization.
								   
\begin{figure}[t]
\begin{center}
\includegraphics[width=6cm]{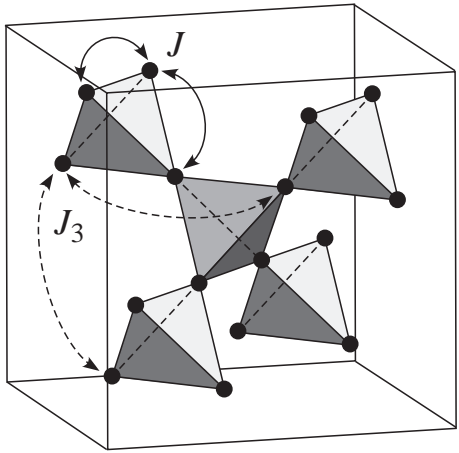}
\end{center}
\caption{
Cubic unit cell of the pyrochlore lattice.
Nearest--neighbor exchanges $J$ and 
third--neighbor exchanges $J_3$ are shown,
which are incorporated in our model for the MC calculations.
See Sec.~2 for the details.
}
\label{fig:pyrochlore}
\end{figure}

Pyrochlore lattice is a typical example of 
the geometrically--frustrated structures;
it consists of a three--dimensional network of corner--sharing tetrahedra
as shown in Fig.~\ref{fig:pyrochlore}.
Pyrochlore lattice is highly frustrated: 
Classical antiferromagnetic Heisenberg models 
with only nearest--neighbor interactions do not show 
any long--range ordering down to the lowest temperature $T=0$ 
\cite{Reimers1991,Moessner1998}. 
The situation is similar in the $S=1/2$ quantum spin model as well
\cite{Canals2000,Harris1991}.
Effects of several perturbations on the highly--degenerate states
have been discussed, such as 
quantum fluctuations
\cite{Harris1991,Tsunetsugu2001},
further--neighbor interactions
\cite{Reimers1991,Reimers1992},
spin--lattice coupling
\cite{Yamashita2000,Tchernyshyov2002}
and Dzyaloshinsky--Moriya interaction
\cite{Elhajal2005}.

Pyrochlore magents are found in many real compounds.
Most typically, so--called $B$ spinel oxides $AB_2$O$_4$
have the pyrochlore structure of the magnetic $B$ cations.
Among the $B$ spinels, chromium spinel oxides $A$Cr$_2$O$_4$
with nonmagnetic $A$ cations 
have attracted much interests recently
\cite{Lee2000}.
In these compounds, each Cr$^{3+}$ cation has
three $d$ electrons in three--fold $t_{2g}$ levels,
which constitute the $S=3/2$ localized spin
by the Hund's--rule coupling.
Hence, $t_{2g}$ levels are half--filled and
there is no orbital degree of freedom
in contrast to the compounds with $B$=V
\cite{Tsunetsugu2003,Tchernyshyov2004}
or Ti
\cite{Matteo2004}.
Therefore, the Cr spinels provide simple $S=3/2$ spin systems
where detailed comparisons between experimental and theoretical studies
can be made rather straightforwardly.

Recently, the chromium spinel oxides have been studied
by applying external magnetic field
\cite{Ueda2005}.
Surprisingly, it was revealed that 
some of them exhibit a half--magnetization plateau
in a very wide range of the magnetic field.
For instance, the magnetization process of HgCr$_2$O$_4$ 
shows a plateau from 10 Tesla to 27 Tesla.

The spin--lattice coupling as a possible mechanism of 
the plateau formation in the Cr spinels 
has been proposed and studied theoretically 
\cite{Penc2004}.
The spin--lattice coupling leads to 
effective biquadratic interaction between spins. 
The $T=0$ study of the classical Heisenberg model 
on the pyrochlore lattice has revealed that 
for finite biquadratic interactions, the model exhibits
a collinear state at intermediate magnetic--field regime, 
in which each tetrahedron has 3--up and 1--down spin configuration.
This collinear state gives rise to a stable half--magnetization plateau.
The magnetization process is favorably compared with
the experimental results.

Main purpose of the present study is to investigate the plateau problem
in the pyrochlore Heisenberg model at finite temperatures
in order to make further comparisons with the experimental results.
We will employ Monte Carlo (MC) calculations
to clarify effects of thermal fluctuations,
which are neglected in the previous $T=0$ study and
may play a crucial role in the frustrated systems.
We will show the temperature dependences of 
the magnetization curve and the magnetic susceptibility
under the external magnetic field, and
summarize the finite--temperature phase diagram.
We will compare the results with the experimental data.

This paper is organized as follows.
In Sec.~\ref{sec2}, we introduce the bilinear--biquadratic Heisenberg model.
MC method is also described.
In Sec.~\ref{sec3}, we show numerical results 
in comparison with the previous $T=0$ results and the experimental data.
Section~\ref{sec4} is devoted to summary and concluding remarks.

\section{Model and Method}\label{sec2}

Our starting point is a classical Heisenberg model
with the spin--lattice coupling of spin--Peierls type
on the pyrochlore lattice under the external magnetic field
\cite{Penc2004}.
The Hamiltonian is given in the form
\begin{eqnarray}
H &=& \sum_{ij} \Big[
J_{ij} (1 - \alpha \rho_{ij}) \mib{S}_i \cdot \mib{S}_j 
+ \frac{K_{ij}}{2} \rho_{ij}^2 \Big] 
\nonumber \\
&-& \sum_i \mib{h} \cdot \mib{S}_i,
\label{eq:HsP}
\end{eqnarray}
where 
$J_{ij}$ is the exchange constant for the spins at the sites $i$ and $j$,
$\mib{S}$ is a classical vector with the modulus $|\mib{S}|=1$,
$\alpha$ is the spin--lattice coupling constant,
$\rho_{ij}$ is the change of the bond length 
between $i$th and $j$th sites, 
$K_{ij}$ is the elastic constant for the $ij$ bond, and
$\mib{h}$ is the external magnetic field.
This model has been studied at zero magnetic field
to describe the low--temperature properties
of Cr spinel compounds
\cite{Tchernyshyov2002}.

In Eq.~(\ref{eq:HsP}), 
the lattice distortions $\rho_{ij}$ are quadratic, 
leading to Gaussian integrals in the partition function, 
and therefore they can be integrated out. 
In general case, the integration results in 
a complicated effective spin--Hamiltonian
with long--range interactions between spin--bonds. 
However, assuming four--sublattice long--range ordering, 
and nearest--neighbor exchange $J$ and elastic coupling $K$ only 
\cite{Penc2004}, 
the effective Hamiltonian 
for the ordered moments appearing 
in the resultant partition function is equivalent to
\begin{equation}
H = \sum_{\langle ij \rangle} J \big\{ \mib{S}_i \cdot \mib{S}_j
- b (\mib{S}_i \cdot \mib{S}_j)^2 \big\}
- \sum_i \mib{h} \cdot \mib{S}_i, 
\label{eq:H}
\end{equation}
where the summation is taken over the nearest-neighbor sites.
The second term in Eq.~(\ref{eq:H}) is the biquadratic interaction, 
with the coupling constant
$
b = J \alpha^2 / K. 
$
Hence, this biquadratic interaction mimics the spin--lattice coupling.
It favors a collinear spin configuration
as the spin--lattice coupling does since $J b$ is always positive.

In the present study, for simplicity and as a first step, 
we neglect all the complications which arise from the lattice distortions 
apart from the effective bilinear--biquadratic part, and  
consider the model (\ref{eq:H})
extended with the third--neighbor exchange $J_3$
(as shown in Fig.~\ref{fig:pyrochlore}).
In particular, we focus on the ferromagnetic $J_3 < 0$
in the following calculations 
since it stabilizes the simple four--sublattice long--range ordering
with the wave vector $\mib{q} = 0$
\cite{Reimers1991}.  
Hereafter, we will set $J=1$ as an energy unit and 
use the convention of the Boltzmann constant $k_{\rm B} = 1$.

We will investigate thermodynamic properties of the model (\ref{eq:H})
by Monte Carlo (MC) calculations
\cite{MotomeUNPUBLISHED}.
We employ the Metropolis algorithm with local spin updates 
to sample spin configurations.
We typically perform $10^5$ MC samplings for measurements
after $2 \times 10^4$ steps for thermalization.
The measurements are performed in every $N_{\rm int}$--times MC update,
and we typically take $N_{\rm int} = 2$.
Results are divided into five bins to estimate statistical errors
by variance of average values in the bins.
We have checked the convergence 
and ergodicity of the results
by comparing those for different initial spin configurations.
The system sizes in the present work are up to $L=8$,
where $L$ is the linear dimension of the system
measured in the cubic units shown in Fig.~\ref{fig:pyrochlore}, i.e.,
the total number of spins $N$ is given by $L^3 \times 16$.

\section{Results and Discussion}\label{sec3}

\begin{figure}[t]
\begin{center}
\includegraphics[width=7cm]{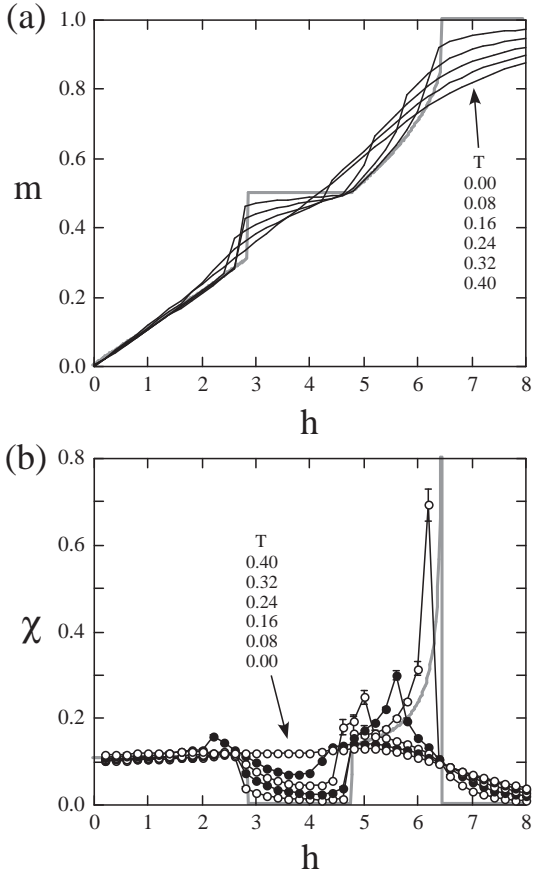}
\end{center}
\caption{
Monte Carlo data at finite temperatures of
the magnetic--field dependences of (a) the magnetization
and (b) the uniform susceptibility 
for $J_3 = -0.05$ and $b=0.1$.
The lines are guides for the eyes.
The bold gray curves show the results at $T=0$
\cite{Penc2004}.
}
\label{fig:m&chi}
\end{figure}

First, we show the magnetic properties of the model.
Figure~\ref{fig:m&chi} shows the magnetic--field dependences of
the magnetization and the uniform magnetic susceptibility
at different temperatures for $J_3 = -0.05$ and $b=0.1$.
The magnetization is defined as
\begin{equation}
m = \langle M_{\rm tot} \rangle 
= \Big\langle \big| \sum_i \mib{S}_i \big| / N \Big\rangle,
\end{equation}
and the susceptibility is measured by
\begin{equation}
\chi = N \big( \langle M_{\rm tot}^2 \rangle 
- \langle M_{\rm tot} \rangle^2 \big) / T,
\end{equation}
where the brackets denote the thermal average.
We show the data for $L=8$ 
since there is no significant system size dependence 
in the range of parameters investigated.

As shown in Fig.~\ref{fig:m&chi}, at low temperatures, 
the magnetization shows a clear plateau feature
at around the half magnetization $m = 1/2$.
At the same time, the susceptibility exhibits a sharp dip
in the corresponding range of the magnetic field.
The results indicate that 
the plateau is stable and survives up to 
$T \sim 0.3J$ for the present parameters.

Next, we examine the phase diagram under the magnetic field.
Following Ref.~\cite{Penc2004},
we measured the order parameters in the form
\begin{equation}
\lambda_\nu = \sum_{\rm tetra} 
\sum_\alpha \Lambda_{\nu,\alpha}^2 / N_{\rm tetra},
\end{equation}
where the first summation is taken over all the 
$N_{\rm tetra} = N/4$ 
independent tetrahedra
(up--pointing ones in the [111] direction).
The index $\nu$ denotes the irreducible representations (irreps)
of the tetrahedral symmetry group $T_d$, 
i.e., $\nu = \{ A_1, E, T_2 \}$,
and the summation on $\alpha$ is taken over members of a given irreps.
The irreps $\Lambda_{\nu,\alpha}$ are given 
by $\rho_{\nu,\alpha}$ in Eq.~(4) in Ref.~\cite{Penc2004}
with replacing $\rho_{i,j}$ by $\mib{S}_i \cdot \mib{S}_j$
on r.h.s of the equation.

\begin{figure}[t]
\begin{center}
\includegraphics[width=7cm]{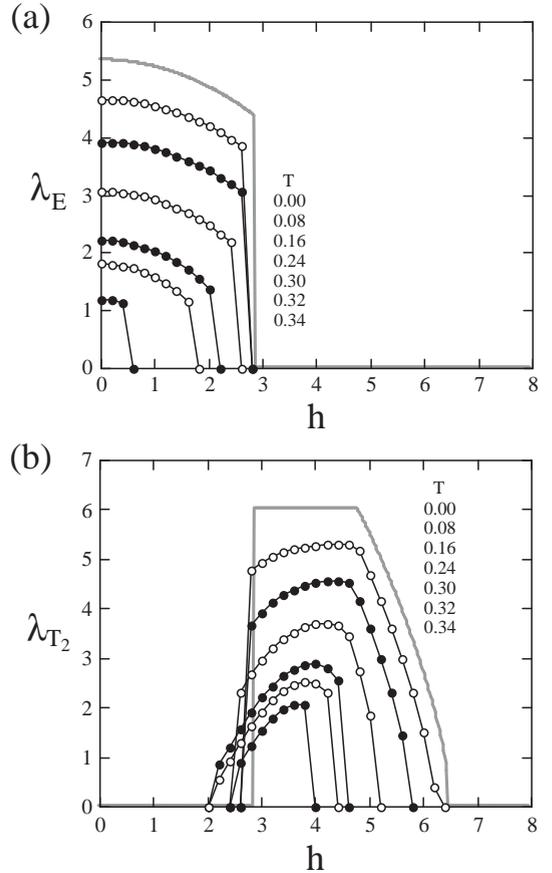}
\end{center}
\caption{
Monte Carlo data of the order parameters with
(a) $E$ symmetry and (b) $T_2$ symmetry
for $J_3 = -0.05$ and $b=0.1$.
The lines are guides for the eyes.
The bold gray curves show the results at $T=0$
\cite{Penc2004}.
}
\label{fig:lambda}
\end{figure}

\begin{figure}[t]
\begin{center}
\includegraphics[width=7cm]{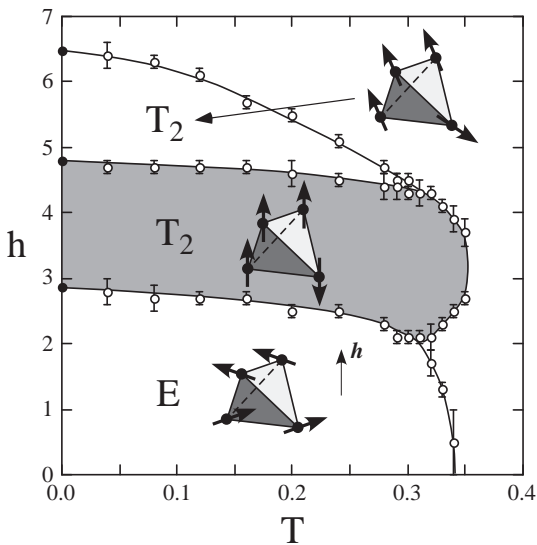}
\end{center}
\caption{
Monte Carlo phase diagram for $J_3 = -0.05$ and $b=0.1$
Symmetry and  typical spin configuration in the tetrahedron
are shown for each phase.
The shaded area denotes the half--magnetization plateau phase.
Filled symbols are the $T=0$ results \cite{Penc2004}.
The curves for the phase boundaries are guides for the eyes.
}
\label{fig:phase}
\end{figure}

Figure~\ref{fig:lambda} shows the MC results for the order parameters
with the symmetry $E$ and $T_2$.
At low temperatures, $\lambda_E$ becomes finite 
in the low--field regime $h \sim 0$.
This corresponds to 
a coplanar long--range ordering with tetragonal symmetry. 
On the other hand, $\lambda_{T_2}$ becomes finite 
in a middle range of the magnetic field $h \sim 4$,
indicating a 3--up 1--down type 
long--range ordering with trigonal symmetry.
Typical spin configurations are shown in Fig.~\ref{fig:phase}.

The $\lambda_{T_2}$ order parameter alone, however, 
does not differentiate between the collinear phase 
which exhibits the magnetization plateau in Fig.~\ref{fig:m&chi} and
the non--collinear but coplanar spin--canting phase
as shown in the phase diagram at $T=0$
\cite{Penc2004}. To identify these phases,
we have examined the collinearity and the coplanarity of the spins
by calculating the so--called nematic order parameters,
and successfully determined the phase boundary between the 
two phases at finite temperatures 
\cite{UNPUBLISHED}.
We note that the obtained collinear--coplanar phase boundary 
corresponds to the shoulder--like feature in the susceptibility 
at $h \sim 4-5$ in Fig.~\ref{fig:m&chi} (b).

Figure~\ref{fig:phase} summarizes the MC phase diagram 
for $J_3=-0.05$ and $b=0.1$,
determined from the data for the system sizes $L=4, 6$ and $8$.
All the phase transitions are of first order, and
the phase boundaries
smoothly 
extrapolate 
the zero--temperature theoretical values.
In addition to the already discussed three ordered phases ---
the $E$ phase, the collinear $T_2$ phase,
the spin--canted $T_2$ phase --- 
at high temperatures there is the disordered paramagnetic phase.
In the figure we also show the 
typical four--sublattice spin configurations on the tetrahedron
for the three ordered phases.
The plateau phase is robust against finite--temperature fluctuations,
and survives up to even slightly higher temperature than 
the critical temperature at zero magnetic field.

The present MC results can be compared favorably with 
the recent experimental results in Cr spinel compounds
such as CdCr$_2$O$_4$ and HgCr$_2$O$_4$.
In particular, 
the magnetization process in the HgCr$_2$O$_4$ 
has been measured up to the saturation field and 
the entire phase diagram has been determined
\cite{Ueda2005}.
Our MC results in Figs.~\ref{fig:m&chi} and \ref{fig:phase}
give a qualitative description of the experimental results.
More quantitative comparison 
including a fine tuning of the model parameters
will be reported elsewhere
\cite{UNPUBLISHED}.

\section{Summary and Concluding Remarks}\label{sec4}

To summarize, 
we have investigated the magnetic properties at finite temperatures
of the antiferromagnetic Heisenberg model
on the pyrochlore lattice by Monte Carlo calculations.
In addition to the ordinary bilinear exchange interaction, 
our model includes the biquadratic interaction as an effective term
describing the coupling of the spins to lattice distortions.
By applying magnetic field,
we have shown that the model exhibits a stable half--magnetization plateau
in a considerable range of temperatures.
The results show good agreement with recent experimental results
in Cr spinel oxides.

Our study indicates that the spin--lattice coupling plays 
a dominant role in stabilizing
the half--magnetization plateau in Cr spinel oxides.
However, we note that other additional elements are indispensable
to understand the properties of real compounds in more detail.
For instance, at zero magnetic field, recent neutron experiments
have found that the spin ordering in the low--temperature phase
is complicated, compound--dependent, and incommensurate
\cite{ChungUNPUBLISHED}. 
To account for such effects, further studies of 
more realistic Hamiltonians are needed, which would include,
e.g., longer--range exchange interactions.

Our further calculations for the model (\ref{eq:H}) 
suggest an interesting phenomenon
when we weaken the strength of the biquadratic interaction $b$.
For small values of $b$, the width (along the $h$ axis) of 
the plateau phase grows with the temperature.
This clearly implies that thermal fluctuations also stabilize the plateau phase.
This tendency becomes conspicuous as $b$ decreases, opening a route toward the
so--called order--from--disorder phenomenon
\cite{Villain1977}.
This interesting issue will be discussed elsewhere
\cite{UNPUBLISHED2}.

Finally, our model is defined for classical spins
although the real compounds have $S=3/2$ quantum spins.
We have also studied the quantum spin problem by spin--wave analysis,
and found that, as expected, the plateau state has a finite spin gap.
The details of the spin excitation spectrum will also be discussed elsewhere \cite{UNPUBLISHED3}.




\section*{Acknowledgment}

The authors acknowledge H. Takagi, H. Ueda, I. Solovyev and M. Zhitomirsky 
for fruitful discussions and useful comments.
Y. M. thanks S.--H. Lee and J.--H. Chung for valuable discussions.
This work is supported by a Grant--in--Aid for Scientific Research
(No. 16GS50219) and NAREGI 
from the Ministry of Education, Science, Sports, and Culture of Japan, 
the Hungarian OTKA Grant T038162 and T049607, and the JSPS-HAS joint project.

\end{document}